\documentclass{aastex}
\usepackage{emulateapj5,apjfonts,psfig}

\usepackage{psfig,graphicx,multicol,color}

\slugcomment{Accepted for publication in The Astrophysical Journal}

\def\cm{{\rm\thinspace cm}}

\def\keV{{\rm\thinspace keV}}

\def\Lsun{\hbox{$\rm\thinspace L_{\odot}$}}

\def\Msun{\hbox{$\rm\thinspace M_{\odot}$}}

\def\ph{{\rm\thinspace ph}}
\def\s{{\rm\thinspace s}}

\def\powerlawfluxat1kev{\hbox{$\ph\cm^{-2}\s^{-1}\keV^{-1}$}}

\def\lapp{\ifmmode\stackrel{<}{_{\sim}}\else$\stackrel{<}{_{\sim}}$\fi}

\def\gapp{\ifmmode\stackrel{>}{_{\sim}}\else$\stackrel{>}{_{\sim}}$\fi}
\def\spose#1{\hbox to 0pt{#1\hss}}

\def\approxlt{\mathrel{\spose{\lower 3pt\hbox{$\sim$}}
        \raise 2.0pt\hbox{$<$}}}
\def\approxgt{\mathrel{\spose{\lower 3pt\hbox{$\sim$}}
        \raise 2.0pt\hbox{$>$}}}

\def\lapp{\ifmmode\stackrel{<}{_{\sim}}\else$\stackrel{<}{_{\sim}}$\fi}
\def\gapp{\ifmmode\stackrel{>}{_{\sim}}\else$\stackrel{>}{_{\sim}}$\fi}

\def\mcg6{MCG$-$6-30-15}
\def\mrk766{Markarian~766}
\def\mr2251{MRC~2251-178}
\def\ngc2110{NGC~2110}
\def\iras13349{IRAS~13349+2438}
\def\iras18325{IRAS~18325--5926}
\def\grs1915{GRS~1915+105}
\def\xtej1748{XTE~J1748-288}
\def\chandra{{\it Chandra }}

\def\xtegammamcg6{$\Gamma=1.9$}

\def\fe25{Fe~{\sc xxv}\,}
\def\fe26{Fe~{\sc xxvi}\,}
\def\Ne9{Ne~{\sc ix }\,}
\def\ne10{Ne~{\sc x }\,}
\def\mg11{Mg~{\sc xi }\,}
\def\si13{Si~{\sc xiii }\,}

\def\apj{ApJ}

\def\apj{ApJ}
\def\apjs{ApJS}
\def\aap{A\&A}
\def\araa{ARAA}

\def\si4{Si~{\sc iv}}
\def\fe25{Fe~{\sc xxv}}
\def\fe26{Fe~{\sc xxvi}}
\def\mg2{Mg~{\sc ii}}
\def\Msun{\ifmmode M_{\odot} \else $M_{\odot}$\fi}
\def\Lsun{\ifmmode L_{\odot} \else $L_{\odot}$\fi}

\tighten

\begin{document}

\shortauthors{Lee \& Ravel}
\shorttitle{Measuring XAFS with Astro~E2}


\title{Determining the grain composition of the interstellar medium
with high resolution X-ray spectroscopy}

\author{Julia C. Lee
\altaffilmark{1,2,3} and
B. Ravel \altaffilmark{4}
}

\altaffiltext{1}{Chandra Fellow; jclee@cfa.harvard.edu}
\altaffiltext{2}{Harvard-Smithsonian Center for Astrophysics, 60 Garden Street MS~4, Cambridge, MA 02138}
\altaffiltext{3}{Massachusetts Institute of Technology, Center for Space Research, 77 Massachusetts Ave., Cambridge, MA. 02139}
\altaffiltext{4}{Center for Corrosion Chemistry and Engineering,  Code 6134,  Naval Research Laboratory, Washington DC 20375}

\begin{abstract}
We investigate the ability of high resolution X-ray spectroscopy to
directly probe the grain composition of the interstellar medium.
Using iron $K$-edge experimental data of likely ISM dust candidates
taken at the National Synchrotron Light Source at Brookhaven National
Laboratory and the Advanced Photon Source at Argonne National
Laboratory, we explore the prospects for determining the chemical
composition of astrophysical dust and discuss a technique for doing
so.  Focusing on the capabilities of the Astro~E2~XRS
micro-calorimeters, we assess the limiting effects of spectral
resolution and noise for detecting significant X-ray absorption fine
structure signal in astrophysical environments containing dust.  We
find that given sufficient signal, the resolution of the XRS will
allow us to definitively distinguish gas from dust phase absorption,
and certain chemical compositions.

\end{abstract}

\section{Introduction}
\label{sec:intro}
UV and IR studies have shown that the heavier elements such as C, N,
O, Mg, Si, and Fe commonly condense to form solid particles
\citep[][and references therein for
reviews]{draineism:review,li_dustreview03,whittet:ism}.  Combined with
planetary studies of meteorites and dust, common forms of interstellar
grains include ice ($\rm H_2O$), graphite~(C), silicates~($\rm SiO_2$;
ferrosilicate, $\rm FeSiO_3$; fayalite, $\rm Fe_2SiO_4$; enstatite,
$\rm MgSiO_3$; and olivine $\rm Mg_2SiO_4$), and various iron species
(pure Fe; w\"ustite, FeO; hematite, $\rm Fe_2O_3$; and magnetite, $\rm
Fe_3O_4$).  UV observational evidence that a fraction of the heavier
elements can be found in grains is \textit{inferred} from the
reduction of certain elements from the gas phase, or the depletion
factor, defined to be the ratio (in gas-phase abundances) of the
amount expected$-$to$-$measured.  The presence of grains is also
\textit{inferred} from studies in the radio and optical. IR emission
spectra, while able to distinguish some complex molecules through
probes of vibrational modes (through excitation of phonons rather than
electrons, as e.g. in the X-rays) are limited by the grain types that
it can sensitively probe (i.e.\ mostly polycyclic aromatic hydrocarbon
or PAHS, graphites, and certain silicates).

Dust particles smaller than 10\,$\mu$m are partially transparent to
X-rays so that the measured absorption in this energy band should be
sensitive to \textit{all atoms in both gas and solid phase}.  High
resolution X-ray spectroscopy therefore provides a powerful and unique
tool for determining the state and composition of the material in the
interstellar medium (ISM).  For interstellar grains, we would detect
oscillatory modulations near the photoelectric absorption edge known
as X-ray absorption fine structure (XAFS).  The XAFS is a condensed
state modulation of the atomic absorption cross-section, thus can
distinguish absorption by the gaseous and the condensed states.  When
a photon excites a deep-core electron into a high-lying, unoccupied
electronic state, the wave function of the outwardly propagating
photoelectron interferes with the portion that scatters from the
surrounding atoms.  This interference produces an oscillatory fine
structure which is characteristic of the chemical species of the
absorber.

The first X-ray study identifying  absorption features with the ISM
were by \cite{mls_crc_ism86} with the Einstein Focal Plane Crystal
Spectrometer pointings of the Crab Nebula.  To date, several ISM
studies using highly absorbed bright X-ray binaries with the Chandra
HETGS and LETGS in the soft ($<1\,\mathrm{keV}$) X-ray have identified
absorption features associated  with ISM material consisting of atomic
and/or molecular forms of oxygen,  neon and iron
\cite[e.g.][]{xrb_paerels:01,
ism_takei:02,schulz_cygx1:02,juett_ism:04}, but with no definitive
indication that the absorption features had to come from grains of
specific chemical makeup.  \cite{jcl_mcg6_wa1:01,jcl_grs1915:02} have
claimed direct detections of iron and silicate grains in Chandra
studies of several black hole systems in the soft X-rays.  In
particular, strong modulations in the form of XAFS seen near the Si~K
edge at $\sim 1.84$\,keV ($\sim$6.7\AA) led \cite{jcl_grs1915:02} to
conclude that they were associated with silicate grains in the
environment of the source or along the line-of-sight.  Unfortunately,
this was reported only as an exciting possibility of a direct
detection of XAFS, as a consequence of the lack of statistics and
because there are $\rm SiO_2$ and polysilicon in the ACIS detectors.
Recently, XAFS detection at Si~K as well as the K-edges of sulfur and
magnesium, attributed to ISM dust have been reported by
\cite{ism_ueda:04} in the spectra of the highly absorbed X-ray
binaries GX~13+1, GX~5-1, and GX~340+0, although Schulz (private
communication) caution that some of these features may be instrumental
in origin.  Chandra X-ray spectra have also revealed detailed edge
structure at $\approxgt 0.7$\,keV near the Fe~L photoelectric edge
which have been attributed to dust (e.g.  \citealt{xrb_paerels:01};
inferred in \citealt{jcl_mcg6_wa1:01}).

While progress is continuing in ISM studies using the Chandra gratings
in the soft X-ray, we consider study of XAFS in the hard X-ray to be a
more promising mechanism for revealing the chemical species of the
condensed state material in the ISM.  Because our better prospects for
X-ray grain studies currently rely on pointings towards bright X-ray
binaries (XRBs) with large amounts of absorption, soft X-ray ISM
studies often have the added worry of needing to separate any solid
state features from the myriad of ionized absorption and/or emission
features originating from the XRB environment (e.g. the accretion disk
corona or the surrounding ionized plasma).  Grain studies in the
$\sim$7.1~keV (1.75\AA) Fe~K regime therefore allow a better search
for XAFS, due to the reduced number of spectral features from atomic
processes in highly ionized plasma occurring in this spectral region.
Furthermore, the chemical shift associated with the oxide species is
more pronounced for the $K$-edge spectra than for $L$-edge spectra,
thus more likely to be resolved using extant instrumentation.  In this
article, we present $K$-edge XAFS data of various iron compounds taken
at the National Synchrotron Light Source (NSLS) at Brookhaven National
Laboratory (BNL) and at the Advanced Photon Source (APS) at Argonne National
Laboratory (ANL).  Using techniques developed for the analysis of
synchrotron data, we discuss the prospect of studying the composition
of iron grains in the ISM with the XRS calorimeter on Astro~E2.  Soft
X-ray XAFS synchrotron studies, and its relevance to studies using
Chandra, XMM-Newton and future X-ray missions will be presented in a
later paper.-- These measurements, together with the FeK measurements
presented in this paper will eventually be jointly analyzed to better
constrain the true crystalline structure of the material.  Ultimately,
combining laboratory and space-based studies of these features as well
as combining hard and soft X-ray measurements to compare all available
absorption edges in the ISM candidate materials will allow us to
investigate the content of the ISM by distinguishing between atomic
and condensed state absorption in the spectra from astrophysical
sources.

\section{The theory behind measuring XAFS}
\label{sec:xafstheoryfrom}

We briefly present theory relevant for the interpretation, processing,
and analysis of XAFS data.  For detailed discussions of XAFS theory
and practice, see e.g.\ \cite{RA-rmp}, \cite{exaftechnique},
\cite{nexafsspec-stohr:96}, and/or \cite{ismdust-krugel:03bk} and
references therein.  See also \cite{astroxafs95}, \cite{astroxafs97},
and \cite{astroxafs98} for early discussions of XAFS in the context of
astrophysical grains.

Measured XAFS spectra for various iron species are shown in the
Appendix, \S~\ref{sec:appendix}.  The sharp rise in the absorption
cross section, also called the edge step, occurs when the incident
photon is equal to the binding energy $E_0$ of the deep-core electron.
The cross section, $\mu(E)$, is typically expressed as
\begin{equation}
  \mu(E) = \mu_0(E) [ 1 + \chi(E) ]
  \label{eq:muchi}
\end{equation}
where $\mu_0(E)$ is the cross section of the isolated atom and
$\chi(E)$ is the oscillatory modulation of the cross section due to
the interaction of the ejected photoelectron with the condensed state
material.  The features within a few electron volts of the edge are
called the near-edge structure.  The extended X-ray absorption fine
structure (EXAFS) extends from a few tens to many hundreds of eV above
the edge.

The oscillatory portion of Eq.~\ref{eq:muchi}, $\chi(E)$~, is
determined by the local-scale atomic structure in the vicinity of
the absorbing atom, thus the absorption spectrum is highly
sensitive to the chemical species of the absorber.  This dependence
is seen in the equation for the oscillatory fine structure in the
single scattering approximation:
\begin{equation}
  \chi(k) = \sum_{i} \frac{S_0^2 F_i(k)}{kR_i^2} \,
  \, e^{-2\sigma_i^2 k^2} \, e^{-2R_i/\lambda(k)}
  \sin[2kR_i + \Phi_i(k)]
  \label{eq:xafs}
\end{equation}
In this equation, the summation is over all atoms $i$ surrounding the
absorber, $F_i(k)$ and $\Phi_i(k)$ are the amplitude and phase of the
function describing the scattering of a photoelectron from neighboring
atom $i$, $R_i$ and $\sigma^2_i$ are the distance between absorber and
scatterer and the thermal RMS variation in that distance, $S_0^2$ is
an element-specific, intrinsic loss term \citep{RA-rmp}, $\lambda(k)$
is the mean free path of the photoelectron, and $k$ --- the wavenumber
of the photoelectron --- is related to the energy by
$k=\sqrt{\frac{2m_e}{\hbar^2}(E-E_0)}$.  The mean free path, the
$1/R^2$ dependence, and the Debye-Waller-like $\sigma^2$ term
together serve to limit the spatial extent probed by the
photoelectron.  XAFS is, therefore, an inherently local probe which
measures structure within 10\,\AA\ or less of the absorbing atom.

As a heuristic explanation of Eq.~(\ref{eq:xafs}), consider the
outwardly propagating photoelectron.  This propagates as a spherical
wave and scatters from the surrounding atoms. (See, for example,
Figure 5 in \citealt{RA-rmp}.)  This backscattered wave interferes
with the outwardly propagating photoelectron.  The density of states
of the absorber is modulated by this interference function measured at
the position of the absorber.  This modulation is, according to
Eq.~(\ref{eq:xafs}), a sum of damped sine waves in photoelectron
wavenumber.

The frequencies of these sine waves are a measure of the distances to
the scatterers, as given by the $\sin(2kR_i)$ term.  The $F_i(k)$ in
the numerator of the pre-factor depends upon the atomic species of the
scatterer.  The oscillatory fine structure function of the EXAFS,
$\chi(k)$ is isolated \citep{AUTOBK} by fitting a smooth spline
representing the atomic contribution to the cross section such that
the low-frequency Fourier components are removed from the spectrum.
The resulting $\chi(k)$ can be Fourier transformed
\citep{exaftechnique} to produce a complex function related to the
partial pair radial distribution function of atoms about the absorber.
Examples of the Fourier transform $\tilde{\chi}(R)$ functions are
shown for various iron compounds in the Appendix, \S\ref{sec:appendix}.

With the aid of theory \citep{feff6,feff8}, the $\tilde{\chi}(R)$
spectra can be analyzed \citep{IFEFFIT} to reveal the details of the
local environment about the absorber, including the atomic species of
and distance to the near neighbors.  Eq.~(\ref{eq:xafs}) is evaluated
for each kind of scatterer in the material.  The contributions from
the scatterers considered in the analysis are summed and then Fourier
transformed.  The Fourier transform of this theoretical $\chi(k)$ is
compared to the Fourier transform of the data.  The parametric terms
in Eq.~(\ref{eq:xafs}), such as $R$ and $\sigma^2$, are modified by
non-linear minimization techniques to best fit the theory to the data
\cite[e.g.][]{newville95:_analy_xafs}.  The Fourier transform is a
complex transform, thus this minimization is made by comparing both
the real and imaginary parts of the transforms of the theoretical and
measured spectra.  A fit to measured data is, therefore, a balance of
contributions from various scatterers which have subtle amplitude
\emph{and} phase relationships.  In \S\ref{sec:appendix}, the
magnitudes of the Fourier transforms of data from several iron
compounds are plotted along with the magnitudes of the Fourier
transforms of the fitted contributions from various near-neighbor
atoms.

Both the near-edge and the extended spectra can be used to identify
the chemical species of the absorber.  The features of the near-edge
structure of metallic and oxidic iron are distinct.  Also the energy
of the onset of the oxide edge is shifted by $\sim10$\,eV relative to
the metal.  In the case of the metallic iron and magnetite comparison
considered in \S\ref{sec:astroe2}, the nearest neighbor in the metal
is another iron atom at a distance of about 2.5\,\AA\ while in
magnetite the nearest neighbor is oxygen at about 1.9\,\AA.  These
differences in distance and species of the near neighbor in the
various iron species are readily apparent in the analysis of the EXAFS
spectra, as seen in the Appendix, \S\ref{sec:appendix}.   For
satellite spectra, we intend to use  similar techniques to identify
the chemical species of grains in the ISM.

\section{Can we measure XAFS with Astro~E2 ? }
\label{sec:astroe2}

In order to investigate the limiting effects of
noise~(\S\ref{subsec:noise}) and spectral
resolution~(\S\ref{subsec:specres}) for detecting an XAFS signal from
satellite measurements, we have made synchrotron XAFS measurements of
various Fe compounds, including metallic Fe, FeO (w\"ustite),
Fe$_2$O$_3$ (hematite), Fe$_3$O$_4$ (magnetite), Fe$_2$SiO$_4$
(fayalite), and FeS$_2$ (pyrite).  In the Appendix
(\S\ref{sec:appendix}; Figs.~\ref{fig:fexafs}) we show the measured
absorption spectra of each of these iron species along with the
spectra convolved by 10 eV to simulate the approximate
resolution\footnote{http://heasarc.gsfc.nasa.gov/docs/astroe/gallery/performance/specres.html}
of the Astro~E2 instrument.  The inset to each figure shows the
Fourier transform of the isolated fine structure spectra.  We see that
the XAFS data contain ample information which can be used to
distinguish different iron species, including the chemical shift of
the onset of the edge, the features in the near-edge structure, and
the details of the Fourier transformed fine structure.  For the
purpose of assessing the feasibility of detailed composition studies
of dust with Astro~E2, we limit our detailed discussion to a
comparison of Fe metal with magnetite Fe$_3$O$_4$.  Our conclusions,
however, are applicable to all iron compounds in the vicinity of the
Fe $K$ edge near 7.1\,keV.

\subsection{The synchrotron experiment}
\label{subsec:synchr-exper}

The data presented in this paper were measured at beamline X11A at
NSLS or at beamline 13BM at the APS.  X11A is a bending magnet
beamline with a double crystal Si(111) monochromator with energy
resolution of about 2x10$^{-4}\,\Delta E/E$.  Adequate harmonic
rejection was accomplished by detuning the second crystal to attenuate
the incident intensity by about 30\%.  13BM was operated in a
substantively similar mode.  The measured samples were a 12\,$\mu$m
iron foil and a fine powder of Fe$_3$O$_4$ dispersed within a quantity
of boron nitride and cold pressed into a pellet.  Both samples were
optimized \citep{size,thickness} for high-quality transmission mode
\citep{exaftechnique} measurement.  The incident and transmitted beam
intensities were measured by ionization chambers filled with
appropriate mixtures of inert gases and operated in a stable voltage
regime.

The synchrotron data were processed using the \textsc{athena} program
\citep{horae}.  Energy calibration was done using a reference Fe foil
measured in parallel during the magnetite measurement.  Normalization
was performed in the standard fashion \citep{exaftechnique}, i.e. by
regressing a two-term polynomial to the region before the pre-edge and
subtracting it from the entire spectrum.  A three-term polynomial was
regressed to a region beyond the absorption edge.  The value of the
post-edge polynomial extrapolated back to the edge energy was used as
the normalization constant.  The \textsc{athena} program was also used
to simulate the effects of noise and energy resolution discussed
below.  Noise was added to the spectra using a pseudo-random number
generator.  The RMS value of the noise is given as a percentage of the
edge step, i.e.\ the normalization constant derived from the
polynomial regressions.  Thus a 10\% level of noise is understood as
having an RMS value of 0.1 on the scale of the normalized spectra
shown in Figs.\ref{fig:xafs-noise} and \ref{fig:xafs-specr}.  The
effect of resolution was simulated by convolving the measured spectra
with a Lorentzian of a given width.

\subsection{The limiting effects of noise}
\label{subsec:noise}

\begin{figure*}
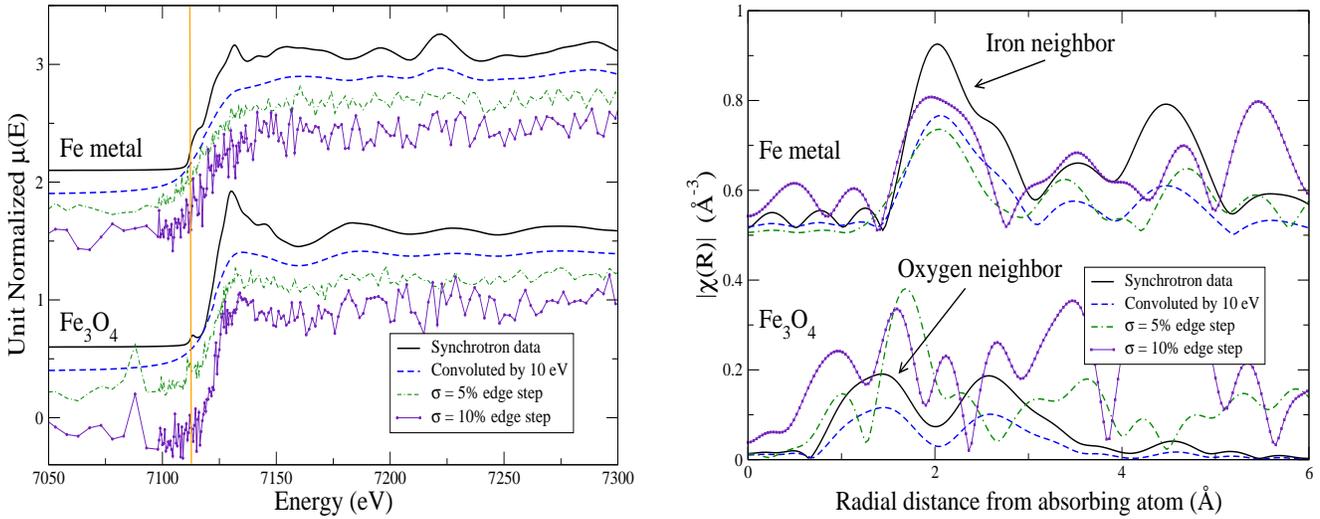

  \hbox{
    \psfig{file=f1a.eps,angle=0,width=0.45\textwidth,height=0.28\textheight}
    \hspace{0.5cm}
    \psfig{file=f1b.eps,angle=0,width=0.45\textwidth,height=0.28\textheight}
    }
  \vspace{0.2in}
  \caption[h]{The effects of instrument resolution and
    noise on synchrotron XAFS spectra of Fe metal and Fe$_3$O$_4$.  In
    both figures, reading from the top, the black (solid) lines shows
    spectra measured at a synchrotron source, the blue (dashed) lines
    show these data convolved by a Lorentzian of 10 eV width to
    simulate the resolution of the Astro~E2 XRS, the subsequent green
    (dash-dot) line shows the convolved spectrum with artificial noise
    of $\rm \sigma_{RMS}$ equal to 5\% of the step height, and the
    bottom purple (dotted) line shows the convolved spectrum with
    $\rm \sigma_{RMS}$ equal to 10\% of the step height.  (LEFT)
    The unit-normalized absorption cross section $\mu(E)$ spectra,
    which emphasize the differences between the metal and the oxide in
    the near-edge portion of the spectra.  The vertical line at
    7112\,eV marks the ionization energy of metallic iron.  For
    gaseous iron, the oscillatory structure blue-ward of the edge step
    seen here will be absent.  (RIGHT) The magnitude of the complex
    Fourier transform of the X-ray fine structure spectra which
    identifies the distance to the neighboring atom.  The first peak
    identifies the position and element of the neighbor atom.  The
    peak just above 2\,\AA\ identifies the iron neighbor in metallic
    iron, whereas the peak below 2\,\AA\ identifies the oxygen
    neighbor in an iron oxide.  Note that convolution and a high level
    of noise leave the iron metal recognizable as such, while the
    noise level is a more serious impediment to identifying the oxide
    species which are more likely to be found in the ISM.}
    \label{fig:xafs-noise}
\end{figure*}

At the spectral resolution of the Astro~E2 XRS, noise will be the most
significant impediment to our ability to speciate interstellar grains.
To demonstrate, we investigate the tolerable noise level by adding
random fluctuations to our synchrotron data at the level of 5--10\% as
described in \S~\ref{subsec:synchr-exper}.  As seen from
Fig.~\ref{fig:xafs-noise}, a S/N~$\approxgt$~10 (or
noise~$\approxlt$10\%) is required to distinguish Fe-rich material in
gas versus the solid state form to enable a direct measurement of the
gas-to-dust ratio in a target absorption column.  At this level, if no
XAFS are detected, we would conclude that the cloud is predominantly
gas phase.

However, should a given environment be predominantly dust, as expected
e.g.\ along the line-of-sight passing through some molecular cloud, we
would like to be able to distinguish the chemical species of the
condensed state of the absorber.  In the comparison of XAFS from iron
metal versus an oxidic form of iron, Fig.~\ref{fig:xafs-noise} shows
that while Fe metal can still be distinguishable at the approximate 
XRS resolution
of 10~eV with a relatively high ($\sigma_{\rm rms} \sim$10\%) level of
noise, a $\sigma_{\rm rms} \sim$5\% noise leaves the oxide
unrecognizable.  Since iron oxides are more likely candidates for ISM
grains, S/N considerations should be one of the most important factors
when estimating adequate exposures times.

Readers should also take caution, when making analogies with
astrophysical environments that the numbers given above are calculated
assuming material in purely solid form.  As such, when considering
astrophysical environments -- the dust contribution to the overall
optical depth of the edge should be accounted for and the numbers
given scaled accordingly.  To illustrate, the required
S/N~$\approxgt$~10 value quoted above would translate to a required
S/N~$\approxgt$~33 (i.e. noise~$\approxlt$3\%) for a pointed
observation of the {\it diffuse} ISM where the dust:to:gas ratio is
$\sim 30$\% \citep[e.g.][]{ism_ref:00}.  Molecular clouds should be
better targets for XAFS studies given the expected higher dust
content.

\subsection{The limiting effects of spectral resolution}
\label{subsec:specres}

\begin{figure*}
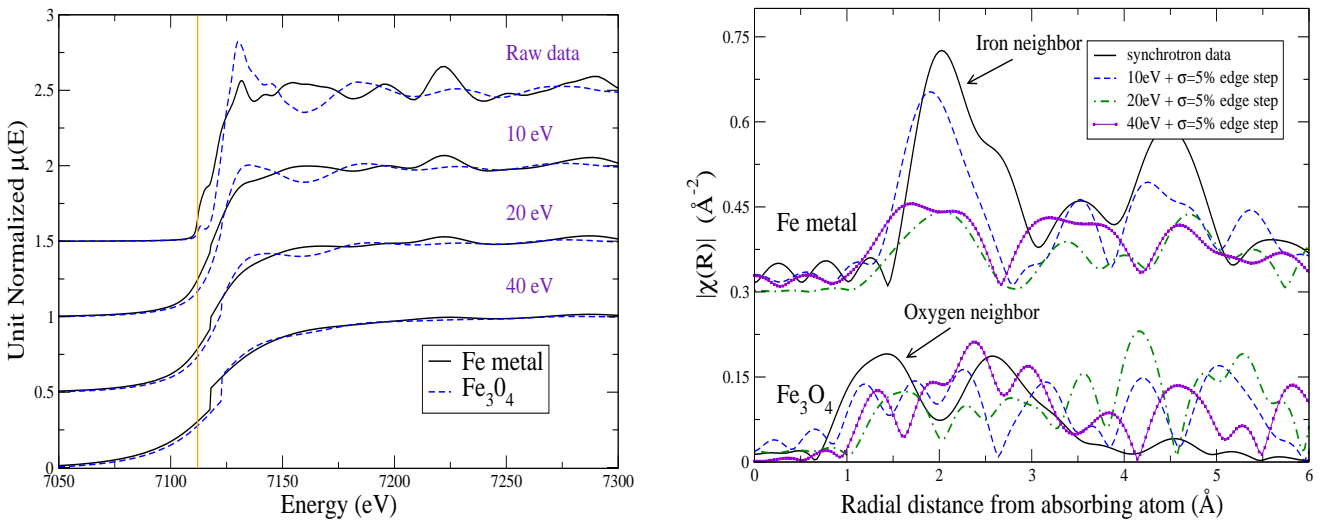

  \hbox{
    \psfig{file=f2a.eps,angle=0,width=0.45\textwidth,height=0.28\textheight}
    \hspace{0.5cm}
    \psfig{file=f2b.eps,angle=0,width=0.45\textwidth,height=0.28\textheight}
    }
  \vspace{0.2in}
  \caption[h]{(LEFT) The effect of convolution on the Fe and Fe$_3$O$_4$
    spectra.  As the width of the Lorentzian convolution is increased,
    the two spectra become increasingly difficult to distinguish.
    (RIGHT) The effect of noise on the Fourier transformed fine
    structure becomes more pronounced as the width of the Lorentzian
    convolution increases.  At 40\,eV convolution width, the peaks
    near 2\,\AA\ identifying the chemical species of the absorber
    become unrecognizable.  Therefore, increasing the width of the
    bins is not a suitable method of improving measurement statistics.
    The full energy resolution of the Astro~E2 XRS along with adequate
    time integration is essential.}
  \label{fig:xafs-specr}
\end{figure*}

We next convolve our synchrotron data of Fe metal and Fe$_3$O$_4$
with Lorentzians to simulate different instrumental resolutions and
find that, to do studies of XAFS at $\sim 7.1$~keV Fe~K, a spectral
resolution of 10\,eV or less (as only currently possible with the
Astro~E2 XRS) is required to extract the X-ray absorption fine
structure information to deduce the chemical make-up of the absorber.
As demonstrated in Fig.~\ref{fig:xafs-specr}, 20~eV resolution with
moderate ($\sigma_{\rm rms} \sim$5\%) noise is still adequate for
identifying metallic iron, but insufficient for distinguishing any
kinds of oxides.  Simulated spectra using the Astro~E2 XRS response
matrix further show that binning the data to increase S/N is
unfortunately not an option.  A doubling of the S/N would require that
we bin the data by a  factor of 4 (i.e. $\sqrt{N}$) implying a 40~eV
resolution -- at this  resolution, even the high S/N synchrotron
spectra of the different compounds become nearly indistinguishable and
vastly more sensitive to noise (Fig.~\ref{fig:xafs-specr} left).

\section{Conclusion}
\label{sec:conclusion}

The unsurpassed spectral resolution of the Astro~E2 XRS in the
$\approxgt$~7.1~keV K-edge region of iron and nickel  will allow the
first detailed study of interstellar grains via analysis of X-ray
Absorption Fine Structure (XAFS). Given data with sufficiently high
$S/N$, the extended portion of the K-edge spectrum can be analyzed to
reveal information about the chemical species of the material in the
ISM.  As demonstrated here, the spectral resolution of Astro~E2 will
allow us to (1) definitively separate dust from gas phase absorption,
and (2) give us good expectations for being able to directly determine
the chemical composition of dust grains containing iron.

We note that, while XAFS theory is quite mature, the study of XAFS is
still largely empirical.  As such, the success of such a study will
require that space-based measurements (e.g. with the Astro-E2 XRS) be
compared with empirical XAFS data taken at synchrotron beamlines to
determine the exact chemical state of the astrophysical dust. Because
determining the content of the ISM has important consequences for
understanding the evolution of the Universe, and X-rays are a powerful
tool for ensuring the success of such a study, \emph{spectral
resolution and area} should remain an important consideration of
future X-ray missions.

\section*{acknowledgements}
We thank M.~Newville for kindly supplying some of the data shown in
this paper.  JCL thanks and acknowledges support from the \chandra\
fellowship grant PF2--30023 $-$ this is issued by the Chandra X-ray
Observatory Center, which is operated by SAO for and on behalf of NASA
under contract NAS8--39073.   JCL is also grateful to Professor Claude
Canizares for providing the freedom and type of atmosphere that
encourages the development of independent ideas throughout her
years  with his group at MIT.


\clearpage


\eject
\vspace{0.2in}

\clearpage

\section{Appendix: XAFS of iron compounds}
\label{sec:appendix}

\begin{figure*}
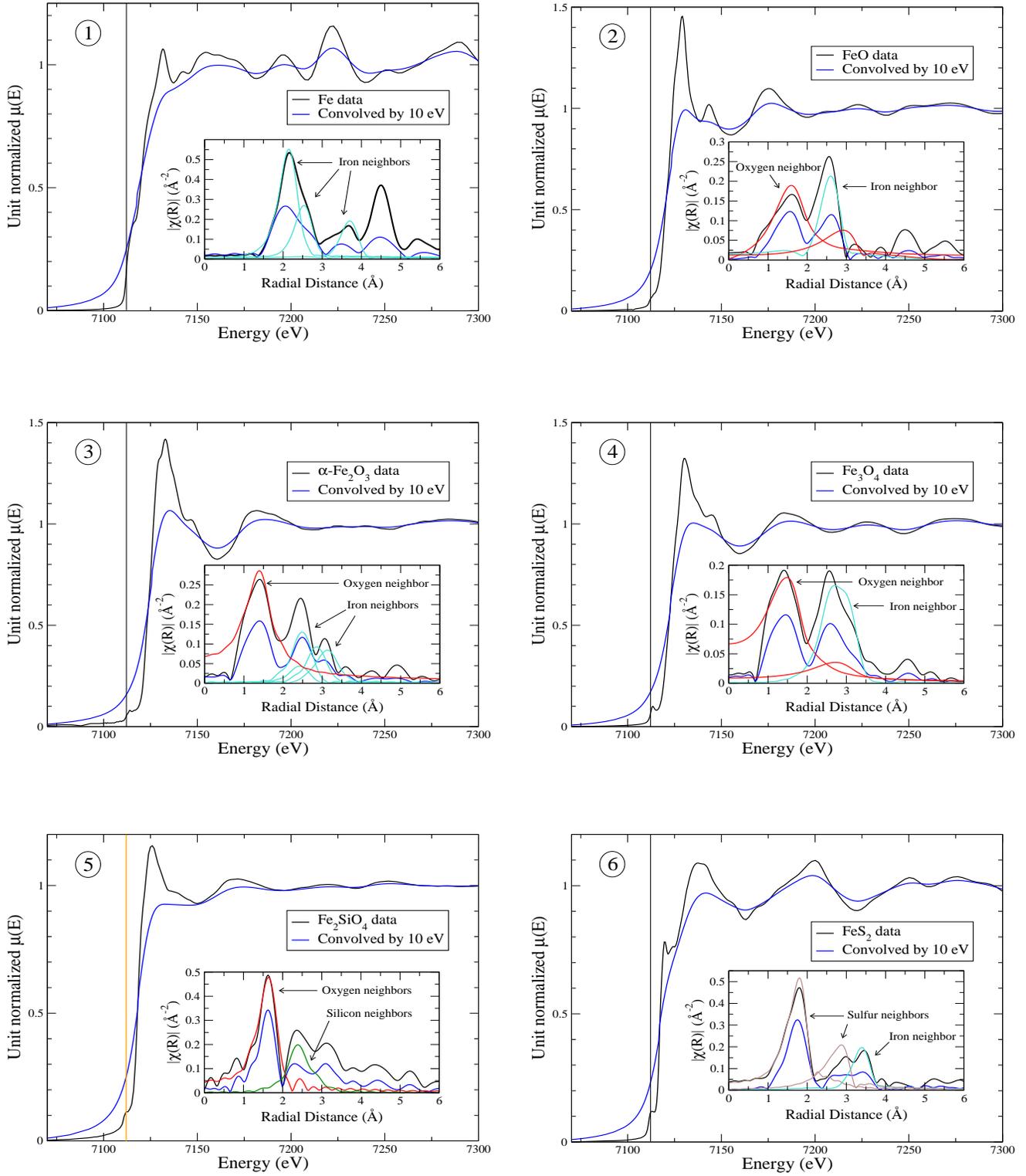

   \hbox{
     \psfig{file=f3a.eps,angle=0,width=0.45\textwidth,height=0.24\textheight}
     \hspace{0.5cm}
     \psfig{file=f3b.eps,angle=0,width=0.45\textwidth,height=0.24\textheight}
     }
   \vspace{0.5in}
   \hbox{
     \psfig{file=f3c.eps,angle=0,width=0.45\textwidth,height=0.24\textheight}
     \hspace{0.5cm}
     \psfig{file=f3d.eps,angle=0,width=0.45\textwidth,height=0.24\textheight}
     }
   \vspace{0.5in}
   \hbox{
     \psfig{file=f3e.eps,angle=0,width=0.45\textwidth,height=0.24\textheight}
     \hspace{0.5cm}
     \psfig{file=f3f.eps,angle=0,width=0.45\textwidth,height=0.24\textheight}
     }
\caption[h]{{The effect of convolution on a variety of common iron
    species: (1) metallic Fe, (2) w\"ustite FeO, (3) hematite
    Fe$_2$O$_3$, (4) magnetite Fe$_3$O$_4$, (5) fayalite
    Fe$_2$SiO$_4$, and (6) pyrite FeS$_2$.  The measured synchrotron
    $\mu(E)$ spectra (black lines) are shown near the edge along with
    those data convolved with a Lorentzian of width 10\,eV (blue
    lines) to simulate the approximate energy resolution of the Astro~E2 XRS.  The
    vertical line in each figure marks the edge energy of metallic
    iron at 7.112\,keV.  Note that the effects of noise are not
    considered here.  The inset of each picture shows the magnitude of
    the complex Fourier transform of the oscillatory fine structure
    after it has been isolated from the $\mu(E)$ as described in the
    text for both the raw (black) and the convolved (blue) data.  Fits
    to the raw data were performed and the contributions from the
    near-neighbor iron (cyan), oxygen (red), sulfur (brown), and/or
    slilicon (green) scatterers are also plotted in the insets.  Note
    the variation in scale of the insets --- the ability to
    distinguish by the details of the fine structure in the presence
    of significant noise is sensitive to the spectral strength of the
    fine structure.}}
\label{fig:fexafs}
\end{figure*}


\begin{thebibliography}{26}
\expandafter\ifx\csname natexlab\endcsname\relax\def\natexlab#1{#1}\fi
\expandafter\ifx\csname url\endcsname\relax
  \def\url#1{{\tt #1}}\fi
\expandafter\ifx\csname urlprefix\endcsname\relax\def\urlprefix{URL }\fi

\bibitem[\protect\astroncite{Ankudinov et~al.}{1998}]{feff8}
Ankudinov, A.~L., Ravel, B., Rehr, J.~J., \& Conradson, S.~D. 1998, Phys. Rev.
  B, 58, 7565

\bibitem[\protect\astroncite{{Draine}}{2003}]{draineism:review}
{Draine}, B.~T. 2003, \araa, 41, 241

\bibitem[\protect\astroncite{{Forrey} et~al.}{1998}]{astroxafs98}
{Forrey}, R.~C., {Woo}, J.~W., \& {Cho}, K. 1998, \apj, 505, 236

\bibitem[\protect\astroncite{{Juett} et~al.}{2004}]{juett_ism:04}
{Juett}, A.~M., {Schulz}, N.~S., \& {Chakrabarty}, D. 2004, \apj,

\bibitem[\protect\astroncite{{Koningsberger} \& {Prins}}{1988}]{exaftechnique}
{Koningsberger}, D.~C. \& {Prins}, R. 1988, {X-ray Absorption: Principles,
  Applications, Techniques of EXAFS, SEXAFS, and XANES; Editors Koningsberger
  \& Prins} (John Wiley \& Sons)

\bibitem[\protect\astroncite{{Kruegel}}{2003}]{ismdust-krugel:03bk}
{Kruegel}, E. 2003, {The physics of interstellar dust} (The physics of
  interstellar dust, by Endrik Kruegel.~IoP Series in astronomy and
  astrophysics, ISBN 0750308613.~Bristol, UK: The Institute of Physics, 2003.)

\bibitem[\protect\astroncite{{Lee} et~al.}{2001}]{jcl_mcg6_wa1:01}
{Lee}, J.~C., {Ogle}, P.~M., {Canizares}, C.~R., {Marshall}, H.~L., {Schulz},
  N.~S., {Morales}, R., {Fabian}, A.~C., \& {Iwasawa}, K. 2001, \apjl, 554, L13

\bibitem[\protect\astroncite{{Lee} et~al.}{2002}]{jcl_grs1915:02}
{Lee}, J.~C., {Reynolds}, C.~S., {Remillard}, R., {Schulz}, N.~S., {Blackman},
  E.~G., \& {Fabian}, A.~C. 2002, \apj, 567, 1102

\bibitem[\protect\astroncite{{Li} \& {Greenberg}}{2003}]{li_dustreview03}
{Li}, A. \& {Greenberg}, J.~M. 2003, in Solid State Astrochemistry, 37--84

\bibitem[\protect\astroncite{Lu \& Stern}{1983}]{size}
Lu, K.~Q. \& Stern, E.~A. 1983, Nuc. Inst. Meth., 212, 475

\bibitem[\protect\astroncite{{Newville} et~al.}{1993}]{AUTOBK}
{M. Newville}, {P. L\=\i vi\c n\v s}, {Y. Yacoby}, {J. J. Rehr}, \& {E. A.
  Stern} 1993, Phys. Rev. B, 47, 14126

\bibitem[\protect\astroncite{Newville}{2001}]{IFEFFIT}
Newville, M. 2001, J. Synchrotron Radiat., 8, 322

\bibitem[\protect\astroncite{Newville et~al.}{1995}]{newville95:_analy_xafs}
Newville, M., Ravel, B., Haskel, D., Rehr, J.~J., Stern, E.~A., \& Yacoby, Y.
  1995, Physica B, 208\&209, 154

\bibitem[\protect\astroncite{{Paerels} et~al.}{2001}]{xrb_paerels:01}
{Paerels}, F., et~al. 2001, \apj, 546, 338

\bibitem[\protect\astroncite{Ravel \& Newville}{2004}]{horae}
Ravel, B. \& Newville, M. 2004, {\textsc{athena} and \textsc{artemis}:
  Interactive graphical data analysis using \textsc{ifeffit}}, to be published
  in Physica Scripta

\bibitem[\protect\astroncite{Rehr \& Albers}{2000}]{RA-rmp}
Rehr, J.~J. \& Albers, R.~C. 2000, Rev. Mod. Phys., 73, 621

\bibitem[\protect\astroncite{{Schattenburg} \&
  {Canizares}}{1986}]{mls_crc_ism86}
{Schattenburg}, M.~L. \& {Canizares}, C.~R. 1986, \apj, 301, 759

\bibitem[\protect\astroncite{{Schulz} et~al.}{2002}]{schulz_cygx1:02}
{Schulz}, N.~S., {Cui}, W., {Canizares}, C.~R., {Marshall}, H.~L., {Lee},
  J.~C., {Miller}, J.~M., \& {Lewin}, W.~H.~G. 2002, \apj, 565, 1141

\bibitem[\protect\astroncite{Stern \& Kim}{1981}]{thickness}
Stern, E.~A. \& Kim, K. 1981, Phys. Rev. B, 23, 3781

\bibitem[\protect\astroncite{{St\"{o}hr}}{1996}]{nexafsspec-stohr:96}
{St\"{o}hr}, J. 1996, {NEXAFS Spectroscopy} (Springer Series in Surface
  Sciences)

\bibitem[\protect\astroncite{{Takei} et~al.}{2002}]{ism_takei:02}
{Takei}, Y., {Fujimoto}, R., {Mitsuda}, K., \& {Onaka}, T. 2002, \apj, 581, 307

\bibitem[\protect\astroncite{{Ueda} et~al.}{2004}]{ism_ueda:04} 
Ueda, Y., Mitsuda, K., Murakami, H., \& Matsushita, K.\ 2004, ArXiv 
Astrophysics e-prints, astro-ph/0410655 

\bibitem[\protect\astroncite{{Whittet}}{2003}]{whittet:ism}
{Whittet}, D.~C.~B. 2003, {Dust in the Galactic Environment (first edition
  1992)} (IOP Publishing Ltd)

\bibitem[\protect\astroncite{{Wilms} et~al.}{2000}]{ism_ref:00}
{Wilms}, J., {Allen}, A., \& {McCray}, R. 2000, \apj, 542, 914

\bibitem[\protect\astroncite{{Woo}}{1995}]{astroxafs95}
{Woo}, J.~W. 1995, \apjl, 447, L129+

\bibitem[\protect\astroncite{{Woo} et~al.}{1997}]{astroxafs97}
{Woo}, J.~W., {Forrey}, R.~C., \& {Cho}, K. 1997, \apj, 477, 235

\bibitem[\protect\astroncite{Zabinsky et~al.}{1995}]{feff6}
Zabinsky, S.~I., Rehr, J.~J., Ankudinov, A., Albers, R.~C., \& Eller, M.~J.
  1995, Phys. Rev. B, 52, 2995

\end{thebibliography}
\end{document}